# Lasing from SOI-integrated GaAsSb nanowires via resonator-driven optical feedback

*J. Zöllner[1*], C. Doganlar[1*], C. García Marcilla[1], S. Meder[1], G. Koblmüller[2] and J. J. Finley[1]*

[1]Walter Schottky Institute, TUM School of Natural Sciences, Technical University of Munich, 85748 Garching, Germany

[2]Institute of Physics and Astronomy, Technical University Berlin, 10623 Berlin, Germany

Silicon photonic integrated circuits critically depend on compact on-chip light sources, for which nanowire (NW) lasers are an attractive solution. However, their practical implementation is often limited by broad emission linewidths and poor frequency stability resulting from weak optical feedback. Here, we integrate individual GaAsSb NWs by transfer-printing onto silicon-on-insulator (SOI) racetrack resonators to realize optical feedback at silicon-transparent wavelengths. Finite-difference-time-domain simulations reveal efficient coupling between the hybrid NW–waveguide mode and the fundamental TE resonator mode, with calculated cavity Q-factors exceeding $10^4$. Experimentally, we observe feedback-induced lasing emission at a low threshold ($P_{th}$) of $8.6 \pm 1.8$ µJ/cm². Compared to identical NW lasers without SOI resonator, the linewidth is reduced by more than a factor of four at $3P_{th}$ and remains stable below 1.8 meV up to $5P_{th}$. Our results demonstrate NW-based light sources on SOI and show that tailored resonator designs enable improved linewidth control and frequency stabilization.

*In this work, both corresponding authors contributed equally.
E-mail: jona.zoellner@tum.de (J.Z.); cem.doganlar@tum.de (C.D.)

Semiconductor nanowires (NWs) have emerged as promising building blocks for next-generation photonic integrated circuits due to their potential for tailoring material composition, band structure engineering and integration. For photonics, the inherent geometry of NWs naturally forms an optical Fabry-Pérot cavity, where the NW itself guides optical modes and end facets serve as mirrors enabling optical feedback and lasing operation[1-7]. At the same time, the nanoscale dimensions provide a small footprint that circumvents strain-induced defect formation when integrated with lattice-mismatched substrates, a critical advantage for monolithic and heterogeneous integration.[8,9] Several previous studies highlighted these beneficial properties, especially for the integration of III-V NW lasers on silicon (Si), or silicon-on-insulator (SOI) photonic waveguides and circuits.[10-15] Furthermore, their emission wavelengths can be precisely tuned through compositional engineering of the active gain medium, offering flexibility for applications in Si photonics.[16-18] Despite these advances, existing NW-laser implementations suffer from inherent limitations that constrain their performance, especially the emission linewidth, stability, modal structure and frequency control. The relatively short cavity lengths (typically 5-10 μm) [10] and moderate end-facet reflectivity result in poor geometrical quality factors (Q-factors) on the order of $10^2$, leading to lasing linewidths in the high GHz range.[20] Additionally, the optical feedback in conventional NW Fabry-Pérot cavities is primarily determined by the refractive index contrast at the NW-air interface, offering limited tunability for threshold optimization, spectral control and integration into photonic integrated circuits.

To mitigate these effects, several concepts were proposed that aim at realizing high-Q resonator cavities either from periodic NW arrays or by embedding single NWs into external cavities. Using inverse design, Mauthe, et al.[21] and Takiguchi, et al.[22] demonstrated 1D photonic crystal (PhC) cavities defined by in-plane NW-arrays, with some aspects of design flexibility that was limited by geometrical constraints. Another strategy is the hybrid integration of single NWs with PhC cavities to realize ultrahigh Q-factors and very small mode volume, an approach that allows access to stable continuous-wave lasing operation and high-speed modulation in Si-based integrated photonic circuits.[23,24] Recently, III-V NW lasers were also integrated in-plane onto high-Q SiN microring resonators to modify lasing threshold and linewidth via evanescent interactions.[25] Extending such designs to lasers operating at Si-transparent wavelengths will open very promising concepts for inter-device coupling through waveguide connections and future optical switching investigations in Si photonic integrated circuits.

In this work, we demonstrate a hybrid III-V NW-laser at Si-transparent wavelengths operating through self-driven optical feedback from a SOI racetrack resonator. In this configuration, we utilize GaAsSb-based NWs emitting at ∼1.1 - 1.2 μm,[26] integrated onto the straight arm of a SOI racetrack resonator using a precision pick-and-place transfer printing technique. This hybrid architecture enables the high-Q Si resonator ($Q_{calc.} = 1.3 \times 10^4$) to provide wavelength-selective optical feedback to the NW gain medium, effectively replacing the conventional Fabry-Pérot mechanism with a resonant feedback system that offers enhanced linewidth control. Thereby, we further demonstrate strong reduction and stabilization in lasing linewidth under high excitation densities. To elucidate the underlying coupling mechanisms and feedback dynamics of this hybrid system, we complement the experiments with a comprehensive electromagnetic simulation study. Using two-dimensional mode analysis and three-dimensional FDTD simulations, we quantify the hybrid mode profiles, coupling efficiencies, and confinement factors governing NW–waveguide interaction, and directly relate these parameters to the experimentally observed spectral characteristics. This combined approach provides a detailed understanding of how the resonator geometry and nanowire positioning enable efficient feedback and waveguide-coupled emission in the SOI platform. **Figure 1** presents the core concept and experimental demonstration of our resonator-enhanced NW emission. The device architecture in **Figure 1**a illustrates our comparative approach: a GaAsSb nanowire (NW1) integrated onto the straight arm of a silicon racetrack resonator, and a nominally identical control nanowire (NW2) placed on a simple Si ridge waveguide.

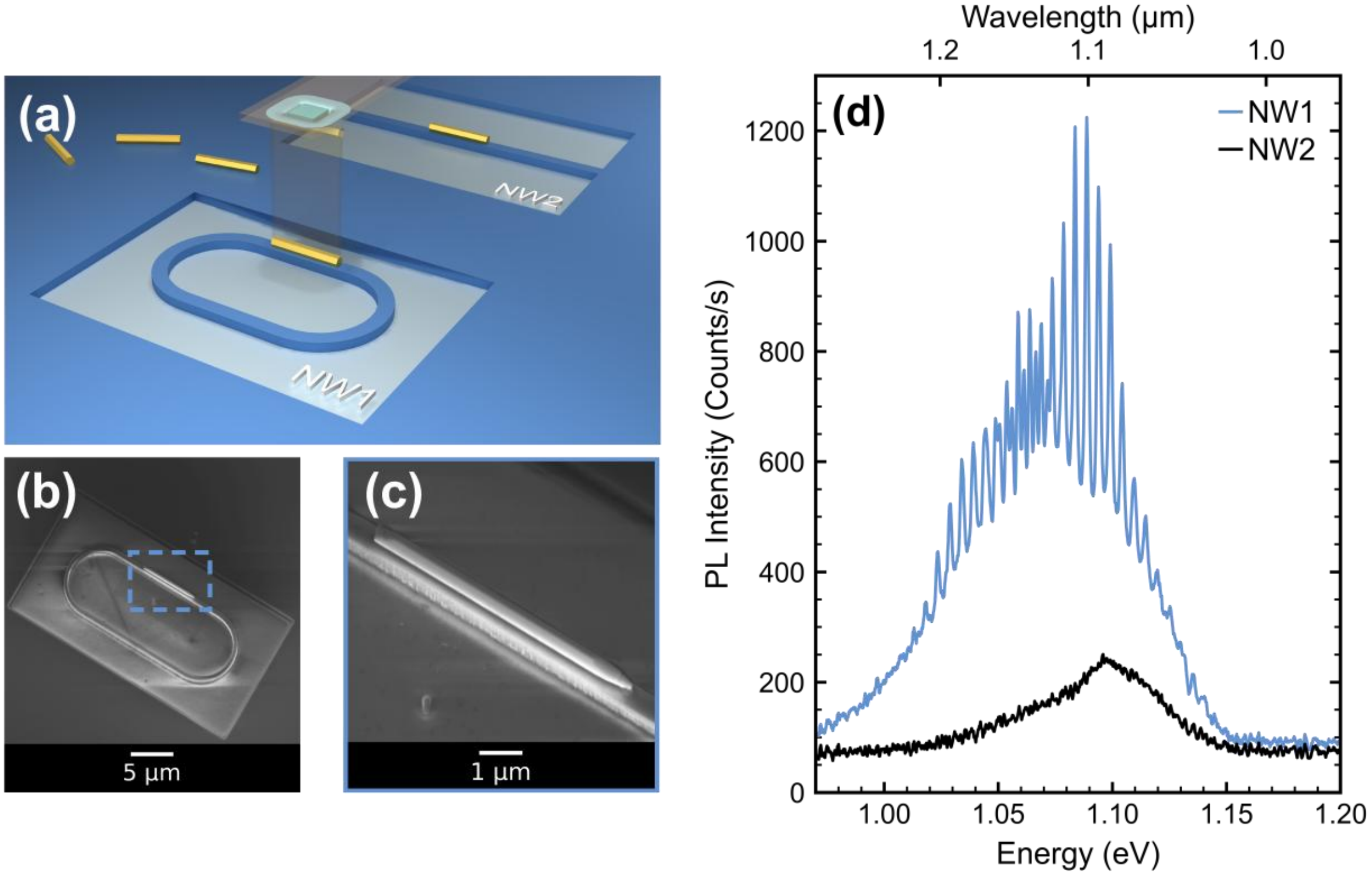

**Figure 1. Hybrid nanowire-resonator platform demonstrating self-driven optical feedback.** (a) Schematic representation showing GaAsSb NW integration on two different Si photonic structures: NW1 positioned on a racetrack resonator arm and NW2 on a reference ridge waveguide. (b, c) SEM images of the fabricated racetrack resonator with integrated NW1, where (c) shows a magnified view demonstrating precise nanowire positioning on the Si waveguide. (d) Photoluminescence spectra comparison at a fixed pump fluence of 35 µJ/cm² revealing distinct spectral modulation in NW1 (blue) characteristic of resonator feedback, in contrast to the emission spectrum of NW2 (black) on the reference waveguide.

The racetrack resonator features 10 μm straight coupling sections with 5 μm radius semi-circular bends, fabricated on a standard SOI platform with 260 nm thick and 500 nm wide Si device layer on a 1 μm thick $SiO_2$ buried oxide (BOX) layer. The two configurations enable direct comparison between resonator-coupled and unmodulated NW emission, emphasizing the effect of the high-Q Si resonator feedback. Figures 1(b) and 1(c) show scanning electron microscopy (SEM) images of the fabricated device after NW integration via precision pick-and-place transfer in our custom-built transfer printing setup (see Supporting Information S1).[27,28] As reported in literature, such heterogeneous pick-and-place integration has recently emerged as a versatile method for creating hybrid NW-architectures that decouple growth optimization processes of NWs from the integration process.[29,30] The overview in **Figure 1**b captures the complete racetrack resonator structure with the integrated NW, while the magnified view in **Figure 1**c confirms the precise alignment of the NW along the Si waveguide. The hexagonal NW cross-section, centered on the 500 nm wide Si waveguide with a lateral positioning accuracy of less than 50 nm, enables efficient coupling between the NW emission and the guided modes of the resonator.

The key experimental result is presented in **Figure 1**d, which compares the photoluminescence (PL) spectra of both configurations under identical excitation conditions. All measurements were conducted at a temperature of 10 K using a Ti:sapphire laser tuned to a wavelength of 780 nm, delivering pulses with a duration of 200 fs at a repetition rate of 82 MHz. The excitation beam was focused onto the NWs with a spot size of (19.9 ± 1.8) μm to ensure homogeneous illumination along the entire NW length. For both spectra shown in Fig. 1(d), a pump fluence of 35 $\mu J/cm^2$ was applied. The spectrum of NW1 (blue) exhibits pronounced periodic modulations superimposed on the spontaneous emission from the GaAsSb NW, with peak intensities significantly exceeding the background level. These spectral features directly demonstrate the wavelength-selective amplification induced by the racetrack resonator's optical feedback. In stark contrast, NW2 (black) displays a smooth, broad emission profile which is found to be intricately linked to the lack of localized photonic modes accessed by the NW. This comparison unambiguously confirms that the observed spectral modulation originates from the resonant coupling between the NW gain medium and the Si racetrack resonator.

To elucidate the physical mechanisms governing light–matter interaction in the hybrid NW-WG platform, we performed a detailed numerical analysis of the supported optical modes and their characteristics. The coupling mechanisms and feedback dynamics in the hybrid NW-resonator system were studied via two-dimensional COMSOL simulations and three-dimensional finite-difference time-domain (FDTD) simulations, performed using Lumerical. **Figure 2**a illustrates the hybrid modes supported by the NW–WG structure, which closely resemble the well-known confined modes of hexagonal NWs[31] as well as the TE and TM guided modes of the Si ridge waveguide. As a first step, we evaluated the coupling efficiency of a guided hybrid mode transitioning into a waveguide mode (**Figure 2**b – upper panel) as a function of waveguide width which was varied from 300 nm to 900 nm. The variation of the WG dimensions provides insights into the optimum conditions for mode confinement and coupling efficiency. Three coupling configurations were considered: the $HE_{1,1,a}$ hybrid mode coupling to the TM mode, the $HE_{1,1,b}$ mode coupling to the TE mode, and the $TE_{0,1}$ mode coupling to the TE mode. All other combinations exhibit negligible coupling (see Supporting Information S2).

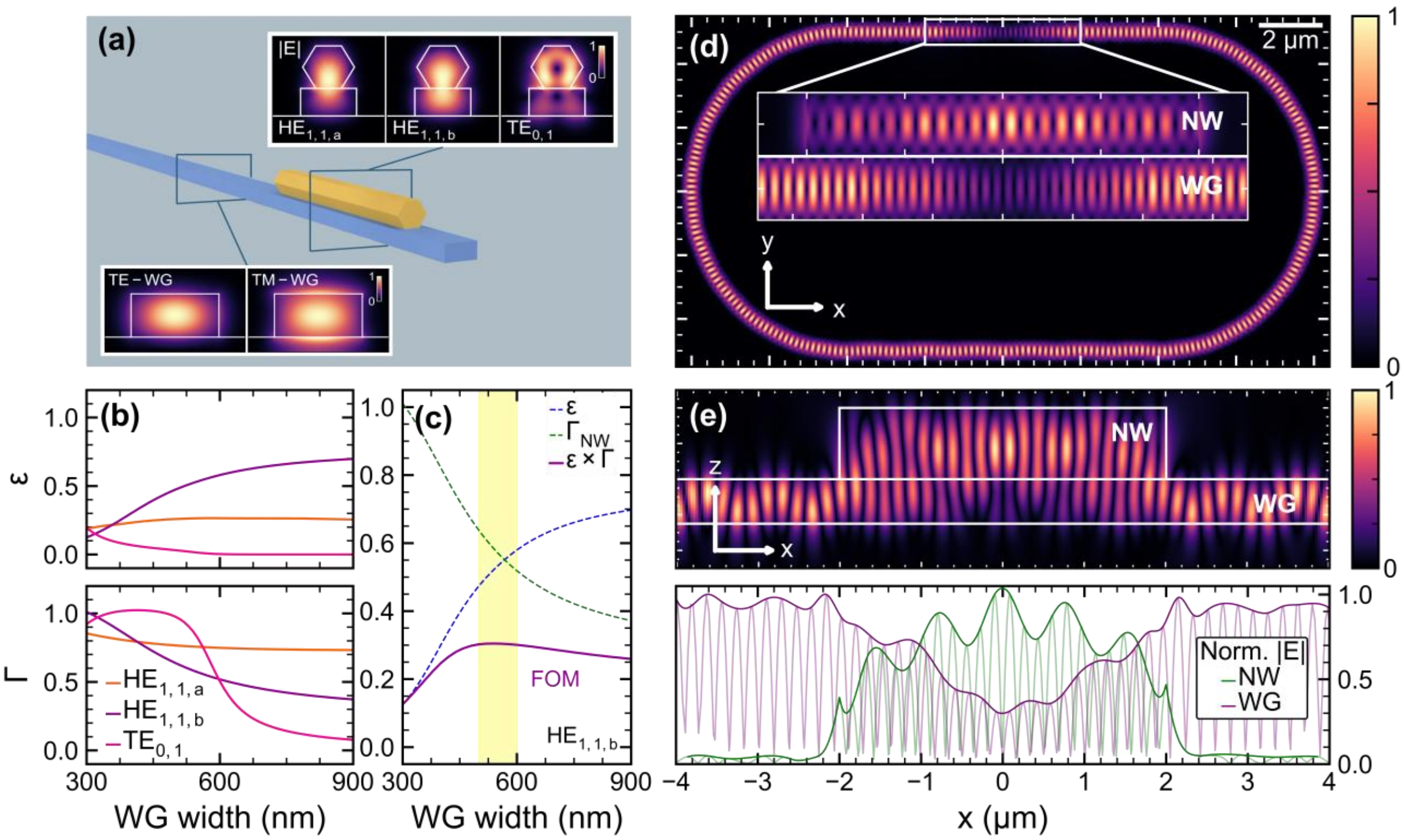

**Figure 2. Nanowire-waveguide coupling simulations** (a) Schematic illustration of the hybrid nanowire–waveguide modes and the confined TE/TM guided modes of the silicon ridge waveguide. (b) Upper panel: Coupling efficiency of hybrid modes as a function of waveguide width for the three mode pairs: $HE_{1,1,a} \rightarrow$ TM, $HE_{1,1,b} \rightarrow$ TE, and $TE_{0,1} \rightarrow$ TE. Lower panel: Confinement factors Γ of the corresponding hybrid modes, illustrating the decreasing overlap with the NW gain medium for increasing waveguide widths. (c) Figure of merit (FOM = ε × Γ) combining coupling efficiency and confinement, with a maximum for the $HE_{1,1,b} \rightarrow$ TE configuration at a waveguide width of 500 nm. (d) 3D FDTD simulation of the electric field intensity distribution in the x,y cross-section through the racetrack resonator for the $HE_{1,1,b}$ excitation. Insets: field distributions for NW and WG cross section. (e) Upper panel: electric field intensity distribution in the y,z cross-section through the NW-WG coupling region showing strong confinement in the silicon ring and efficient coupling between the NW and the underlying waveguide. Lower panel: normalized field distribution in NW and WG cross-section demonstrating quantitative coupling strength.

For both hybrid electric modes, the coupling efficiency increases with increasing WG width. In contrast, the $TE_{01}$ mode exhibits reduced coupling as the WG width increases due to the reduced spatial overlap of the polarization components with the TE WG mode (see Supporting Information S2). Furthermore, it is important to note that all hybrid guided modes already contain a non-negligible field fraction in the waveguide, meaning that coupling efficiency alone does not fully capture the performance of the coupled structure. The spatial overlap with the active NW gain medium is equally important, and this is quantified by the confinement factor[32] given by,

$$\Gamma = \frac{2\varepsilon_0 c n_a \iint_{active} dx\, dy\, |\bar{E}|^2}{\iint dx\, dy \left[\bar{E} \times \bar{H}^* + \bar{E}^* \times \bar{H}\right] \cdot \hat{z}}$$

shown in the lower panel of **Figure 2**b with the refractive index of the homogeneous GaAsSb active region $n_a$. Here, the integral in the numerator considers the cross-section of the homogeneous gain medium while the denominator integrates over the entire cross-section of the NW-WG system. For all three modes, Γ decreases with increasing waveguide width, as a larger fraction of the optical power shifts from the NW into the wider waveguide region. As an optimal device requires both strong confinement in the gain medium and efficient coupling into the Si waveguide, we introduce a figure of merit FOM = ε × Γ, shown in **Figure 2**c. The FOM reaches its maximum for the $HE_{1,1,b}$ hybrid mode coupling to the TE waveguide mode at a waveguide width of approximately 500 nm. This justifies the choice of our experimental WG width used throughout this study.

This result further motivated full 3D FDTD simulations of the proposed racetrack resonator using the $HE_{1,1,b}$ mode as the injected source. Spectral analysis of the coupled system, obtained through Fourier transformation of time-monitored signals within the NW region, reveals multiple discrete resonance peaks (see Supporting Information S3). These sharp resonances correspond to different longitudinal modes of the ring cavity that satisfy the resonance condition. A resonant mode at an energy of 1.12 eV, close to the maximum gain of our GaAsSb NWs, was chosen to calculate the electric field distribution viewed from different perspectives. The simulated in-plane (x-y) electric-field distribution (**Figure 2**d) reveals strong optical confinement in the silicon racetrack resonator with the characteristic standing-wave pattern of the circulating TE mode. The inset cross-section through the coupling region highlights the vertical mode overlap between the NW and the waveguide. The hexagonal GaAsSb NW is positioned directly on top of the Si WG, enabling efficient evanescent coupling between the hybrid NW mode and the fundamental TE waveguide mode of the resonator. The vertical field profile (**Figure 2**e – upper panel) confirms that the electric field extends continuously from the waveguide into the NW, demonstrating strong penetration into the active region which is quantified in the lower panel of **Figure 2**e. Here we find that more than 50% of the TE WG mode is transferred to the NW in the center of the coupling region. The correlation between coupling efficiency, confinement factor, and spatial field distributions validates our design choices and confirms the self-driven optical feedback scheme. Recent work from Yi et al. has demonstrated microring-assisted NW lasing on a silicon nitride photonics platform at room temperature[25], highlighting the general potential of resonator-mediated feedback for stabilizing NW-type emitters. In contrast, our approach establishes a hybrid GaAsSb NW laser directly integrated on the SOI platform, where the NW couples to the fundamental TE mode of a Si waveguide resonator operating in the Si-transparent near-infrared regime. Importantly, the NW is positioned on top of a pre-fabricated photonic component, enabling a transfer-based integration approach that supports post-fabrication assembly on Si photonic structures in a very flexible manner. With the higher refractive index of Si as compared to SiN, our device offers a footprint as small as 10 µm × 15 µm, which is comparable to state-of-the-art WG coupled NW-induced PhC cavity designs.[23]

Having established the geometric configuration and optical mode engineering underlying the efficient NW–resonator coupling, we now present the experimental findings of this hybrid integration under optical pumping. In particular, we investigate how the strong evanescent interaction and self-driven optical feedback discussed in relation to **Figure 2** is displayed in the emission characteristics of individual GaAsSb NWs emitting at Si-transparent wavelengths. By directly comparing the pump-fluence-dependent micro-photoluminescence (µ-PL) response of NWs coupled to a racetrack resonator with that of reference standalone NWs lying on sapphire, we isolate the role of resonator-induced feedback shaping the spectral characteristics.

**Figure 3** illustrates this direct comparison of the two different integration cases, highlighting the impact of resonator-induced optical feedback combined with efficient waveguide coupling. First, **Figure 3**a depicts the recorded spectra of a single GaAsSb NW integrated on a racetrack resonator under pulsed excitation at a temperature of 10 K. At low excitation levels, the emission exhibits a smooth and broad spectrum spanning approximately between 1.0 – 1.15 eV (~1.1 - 1.2 µm), consistent with spontaneous emission from bulk GaAsSb NWs.[26] With increasing pump fluence (0.8 µJ/cm²), pronounced spectral modulations start to emerge, indicating the onset of optical feedback induced by the photonic resonator. At elevated pump fluences (> 9.1 µJ/cm²), equally spaced peaks dominate the emission spectrum, arising from efficient coupling of the NW emission into the propagating modes of the racetrack resonator. The dominant mode is highlighted by a purple arrow. The spacing between adjacent resonances ($\Delta E$ = 5.2 meV) is characteristic of the extended optical path length ($L_{RR}$ ~ 51.4 µm) of the racetrack resonator as compared to the standalone NW ($L_{NW}$ ~ 5 µm), confirming that they originate from the coupled system. This observation demonstrates the anticipated optical feedback, validating the central design goal of the integrated device. The inset shows a microscope image of the device under laser excitation, where both the emitted light from the NW and the scattered light coupled to the resonator are visible. The input–output characteristics of the dominant resonant peak are shown in **Figure 3**b. The peak intensity (purple) is plotted as a function of pump fluence alongside the averaged intensity of the two adjacent spectral minima (blue). A clear nonlinear increase in the peak intensity marks the onset (8.6 ± 1.8 µJ/cm²) of stimulated emission. At high pump fluences (> 30 µJ/cm²), a reduction in emission intensity is found, an observation that we attribute to heating effects. The lasing threshold is determined from linear fits to the spontaneous emission regime below threshold (blue line) and the stimulated emission regime above threshold (purple line), with the intersection defining the threshold pump fluence. **Figure 3**c further demonstrates lasing behavior by presenting the input–output characteristics on a log–log scale. The resulting S-shaped curve is characteristic of the transition from spontaneous to stimulated emission. Starting with a linear spontaneous emission regime the extracted slope increases at threshold and subsequently approaches unity at higher pump fluences, consistent with the expected behavior of a lasing system. The less pronounced S-like characteristic in the LL-curve of the resonator device as compared to typical standalone NWs[6, 11] is attributed to strong mode competition and high β-factors (see Supporting Information S3).

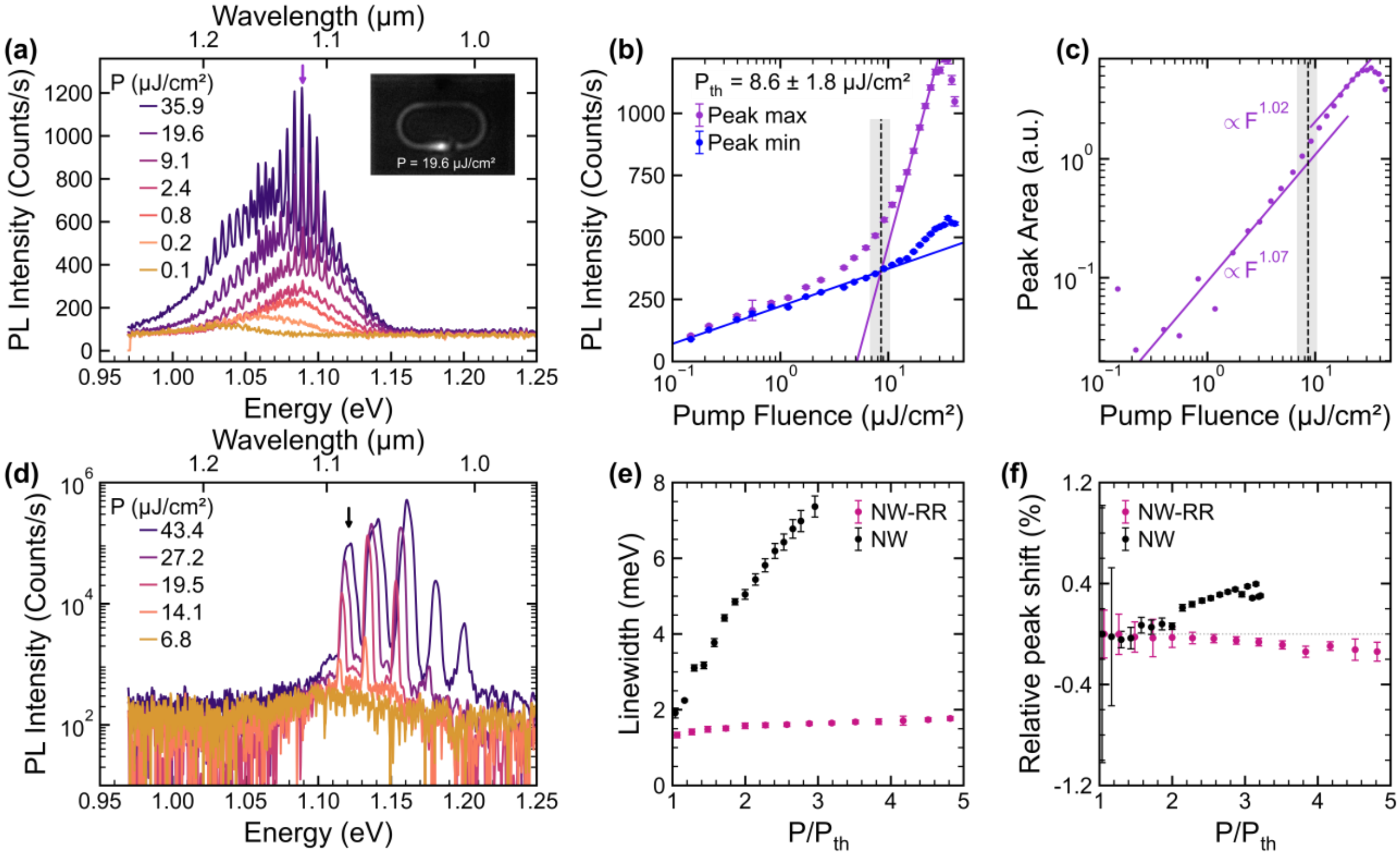

**Figure 3. Comparison between pump-fluence-dependent μ-PL measurements of an individual GaAsSb NW on sapphire versus on a racetrack resonator.** (a) Spectra of a GaAsSb NW coupled to a racetrack resonator showing lasing emission at 10K. The dominant lasing peak is indicated by the purple arrow. Inset: microscope image of optically excited NW-resonator system. (b) Input–output characteristics of the dominant peak (purple) and adjacent spectral minima (blue), with the lasing threshold (dashed line) extracted from linear fits. (c) Log–log input–output characteristics showing the typical S-shaped curve of a laser. (d) Spectra of a standalone GaAsSb NW on sapphire at 10K, where optical feedback arises from NW facets. (e) Linewidth comparison between standalone NW (black) and resonator-coupled NW (magenta) showing reduced linewidth and enhanced stability for the resonator system. (f) Corresponding relative lasing peak shift, normalized to the respective peak energy at $P=P_{th}$, highlighting improved spectral stability for the ring-resonator-coupled device as compared to the standalone NW.

To assess the reproducibility of the observed behavior, we performed pump-fluence-dependent μ-PL measurements on multiple nominally identical GaAsSb NW devices integrated on racetrack resonators and compared them to standalone NWs on sapphire. Across all measured resonator-coupled NWs (see Supporting Information S4), we consistently observe the emergence of resonator-mode modulation at elevated excitation densities, accompanied by comparable lasing thresholds ranging between 8.6 and 15.5 μJ/cm$^2$. Importantly, in all cases the resonator-integrated configuration exhibits enhanced spectral stability at higher pump fluences: across three additional devices the linewidth remains nearly constant at ~2 meV above threshold (see Supporting Information S4). These statistics confirm that the reduced threshold and improved frequency stability are intrinsic features of the ring-resonator-coupled geometry rather than a single-device effect.

For comparison, **Figure 3**d shows pump-fluence dependent μ-PL spectra of an equivalent standalone GaAsSb NW lying on a sapphire substrate under the same excitation conditions. In this configuration, optical feedback is provided solely by reflections at the NW end-facets, resulting in a more than 20 times lower calculated geometrical quality factor as compared to the resonator-based system. Nonetheless, a transition to lasing behavior is observed when increasing pumping. Figures 3e and 3f quantitatively compare the measured linewidths and frequency stability of both systems. As shown in **Figure 3**e, the emission linewidth narrows sharply at threshold, reaching 2.0 meV for the standalone NW (black) and 1.3 meV for the ring-resonator-coupled NW (magenta). At elevated pump fluences, the resonator-integrated system exhibits significantly enhanced spectral stability. At three times the threshold pump fluence, the linewidth increases by only ∼20%, whereas the standalone NW shows more than 300% broadening from 2.0 meV to 7.3 meV under these conditions. Consistently, the improved spectral stability is also evident in **Figure 3**f, which plots the relative shift of the dominant lasing peak above threshold with respect to its spectral position at $P = P_{th}$. At an excitation fluence of $3P_{th}$, the standalone NW on sapphire exhibits a pronounced blue-shift of approximately 0.4%, whereas the resonator-coupled NW shows only a minor red-shift of roughly 0.05%. The larger peak shift observed for the standalone NW is attributed to a carrier-induced decrease of the refractive index[33], as the entire cavity is formed by the active NW material. In contrast, the effective cavity of the resonator-integrated device is largely defined by the passive Si racetrack resonator, which mitigates effects arising from refractive index variations of the active medium and thereby stabilizes the lasing wavelength. The slight red-shift observed in the resonator-coupled case is consistent with residual thermal heating of the device at elevated pump fluences, leading to a temperature-induced increase of the effective refractive index for both the active NW and passive resonator material. Our observations regarding lower lasing thresholds, narrower linewidth and enhanced frequency stability are in good agreement with the recently published findings from Yi et al.[25]

In summary, we demonstrated a hybrid GaAsSb nanowire–silicon racetrack resonator platform that enables resonator-mediated optical feedback and efficient evanescent coupling into guided TE WG modes. The integrated devices exhibit pronounced spectral modulation and the emergence of regularly spaced narrow peaks, evidencing cavity-enhanced amplification and stimulated emission. Lasing is consistently observed across multiple integrated devices with thresholds of 8.6 – 15.5 $\mu J/cm^2$ and linewidths remaining near 2 meV, even well above threshold. The best-performing device exhibits a threshold of 8.6 ± 1.8 $\mu J/cm^2$, a linewidth of 1.3 meV at threshold, and a relative peak shift of less than 0.1% up to $P = 3P_{th}$. Compared to standalone GaAsSb nanowires on sapphire, this platform enables reduced lasing thresholds and strongly improved linewidth and emission-energy stability. These results establish low-threshold lasing from individual GaAsSb NWs directly on a silicon photonics platform, providing a compact and scalable route toward nanoscale near-infrared light sources that are integrable into photonic integrated circuits. Future work will focus on deterministic mode control and cavity engineering to enable robust single-mode operation.

**Acknowledgements:**

The authors sincerely thank H. Riedl for experimental support. This work was supported financially by the Horizon Europe OptoSilicon project (ID: 964191), as well as the Bright Chips (ID: 101166915) and QUANtIC projects (ID: 771747), funded by the European Research Council. We also gratefully acknowledge financial support from the DFG clusters of excellence MCQST (EXC 2111) and e-conversion (EXC 2089).

Supporting Information for

# Lasing from SOI-integrated GaAsSb nanowires via resonator-driven optical feedback

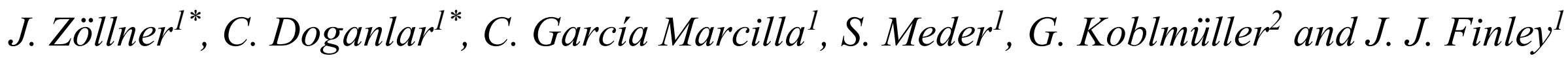

[1]Walter Schottky Institute, TUM School of Natural Sciences, Technical University of Munich, 85748 Garching, Germany

[2]Institute of Physics and Astronomy, Technical University Berlin, 10623 Berlin, Germany

***In this work, both corresponding authors contributed equally.**
E-mail: jona.zoellner@tum.de (J.Z.); cem.doganlar@tum.de (C.D.)

## S1. Fabrication and Transfer Printing

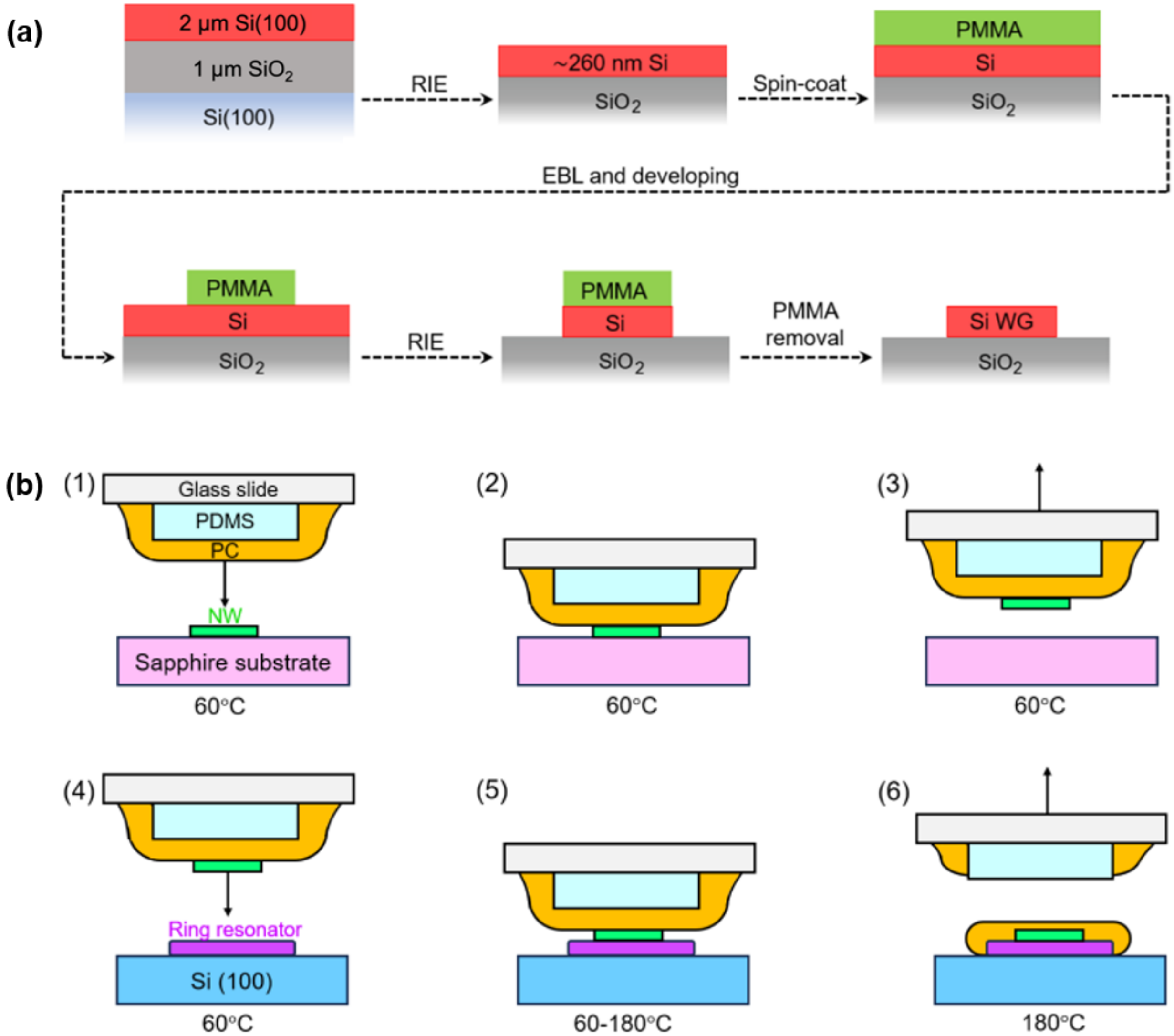

**Figure S1: Fabrication flow diagram of the SOI WG platform and NW transfer printing technique.** (a) Fabrication flow for defining SOI waveguides. The Si device layer is thinned by RIE, patterned using PMMA and EBL, and subsequently etched to form the silicon waveguide structures on $SiO_2$. (b) Schematic of the transfer-printing process used to deterministically place GaAsSb NWs from a sapphire substrate onto the SOI racetrack resonator using a PDMS/PC stamp, followed by thermal release and PC removal.

**Figure S1a** illustrates the main fabrication steps used to prepare the silicon-on-insulator (SOI) racetrack resonator. The process starts from a SOI wafer consisting of a 260 nm thick Si(100) device layer on top of a 1 μm buried $SiO_2$ layer, supported by a thick Si(100) handle wafer. First, a PMMA resist layer is deposited by spin-coating. This resist serves as an electron-beam lithography (EBL) mask, allowing the waveguide pattern to be written with high precision. Following EBL exposure and development, the remaining PMMA defines the waveguide layout. The pattern is then transferred into the silicon layer using reactive ion etching (RIE), selectively removing the exposed silicon while leaving the masked regions intact. Finally, the PMMA is removed, resulting in the final silicon waveguide structures on the $SiO_2$ buried oxide layer. **Figure S1**b illustrates the deterministic transfer-printing procedure used to integrate individual GaAsSb NWs onto the silicon racetrack resonator. The nanowires are initially dispersed on a sapphire substrate (where they were measured for reference) and are subsequently picked up using a transparent stamp consisting of a PDMS support layer coated with a thin polycarbonate (PC) sacrificial film. The pickup step is performed at 60 °C, enabling reliable adhesion of the nanowires to the PC layer. After pickup, the stamp is aligned to the target silicon photonic circuit and brought into contact with the straight arm of the racetrack resonator, again at 60 °C. This allows controlled placement of the nanowire at the desired position on the resonator with sub-50 nm alignment accuracy. Following placement, the sample is heated to 180 °C, i.e., above the glass transition temperature of PC, which enables release of the sacrificial layer from the stamp and leaves the nanowire positioned on the resonator. Finally, the PC layer is removed by dissolution in chloroform. Notably, the nanowires remain attached to the silicon surface and are not displaced during the solvent-based removal process, indicating strong adhesion between the nanowire and the device surface. Importantly, transfer printing does not measurably degrade the optical performance: test measurements performed on the same nanowire before and after transfer show no detectable change in lasing threshold, consistent with previous reports by Jevtics et al.[1]

## S2. Polarization Resolved Mode Profiles

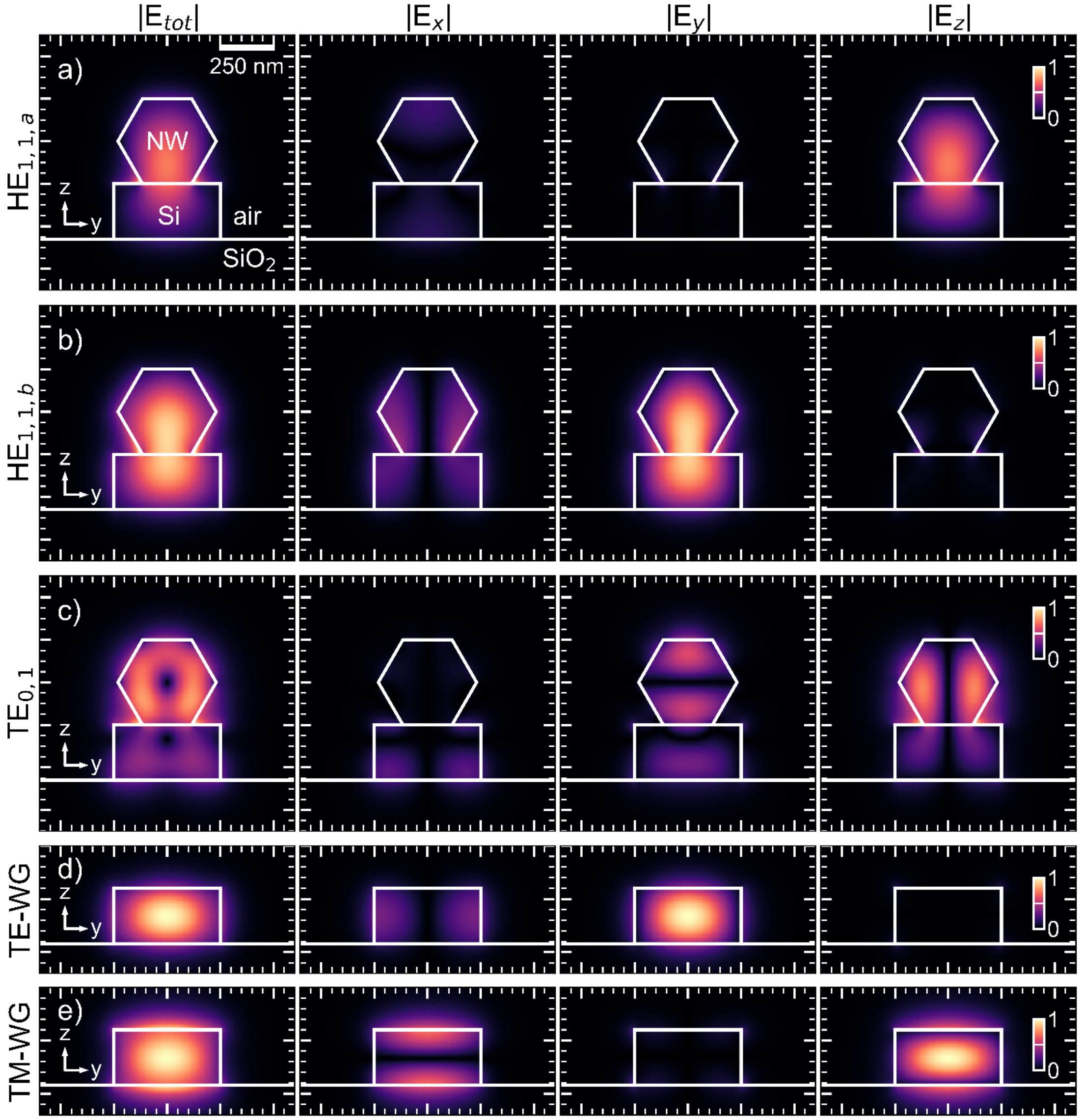

**Figure S2: Simulated guided-mode field distributions of the NW-WG hybrid structure and standalone WG.** (a–c) Calculated electric-field intensity profiles of the lowest-order hybrid eigenmodes supported by the combined hexagonal NW and rectangular SOI waveguide cross section. The white outlines indicate the simulated geometry consisting of the GaAsSb NW on top of the Si WG with $SiO_2$ bottom cladding surrounded by air/vacuum. The scale bar corresponds to 250 nm. Shown are the total field magnitude $|E_{tot}|$ and the individual Cartesian field components $|E_x|$, $|E_y|$, and $|E_z|$. Panels (a) and (b) correspond to two $HE_{1,1}$-like hybrid modes ($HE_{1,1,a}$ and $HE_{1,1,b}$), while (c) shows the $TE_{0,1}$-like mode. (d,e) Reference field distributions of the fundamental TE and TM waveguide modes of the bare SOI waveguide without the NW. All field maps are normalized to their respective $|E_{tot}|$.

**Figure S2** visualizes the simulated optical eigenmodes of the hybrid system formed by the GaAsSb nanowire placed on top of the silicon waveguide. The field maps reveal how the supported modes and their polarization components are spatially distributed, i.e., how strongly the optical field overlaps with the nanowire versus being confined in the SOI waveguide. The two lowest-order modes $HE_{1,1,a}$ and $HE_{1,1,b}$ exhibit strong intensity localization within the NW cross section while simultaneously maintaining appreciable overlap with the underlying silicon waveguide. This confirms that the nanowire does not act as an isolated cavity but becomes optically coupled to the silicon photonic circuit. The $TE_{0,1}$-like mode shows a similar pattern as compared to the $TE_{0,1}$ guided modes typically found in conventional standalone NWs. Finally, the bottom panels show the fundamental TE and TM modes of the bare silicon waveguide, serving as a reference for comparison and highlighting how the nanowire perturbs and hybridizes the guided-wave modes. Overall, the simulations demonstrate that the hybrid structure supports low-order modes with significant overlap between the nanowire gain region and the SOI waveguide mode, enabling efficient coupling of nanowire emission into the silicon resonator circuit.

## S3. FDTD Spectra, Q-factor and Beta-factor Calculations

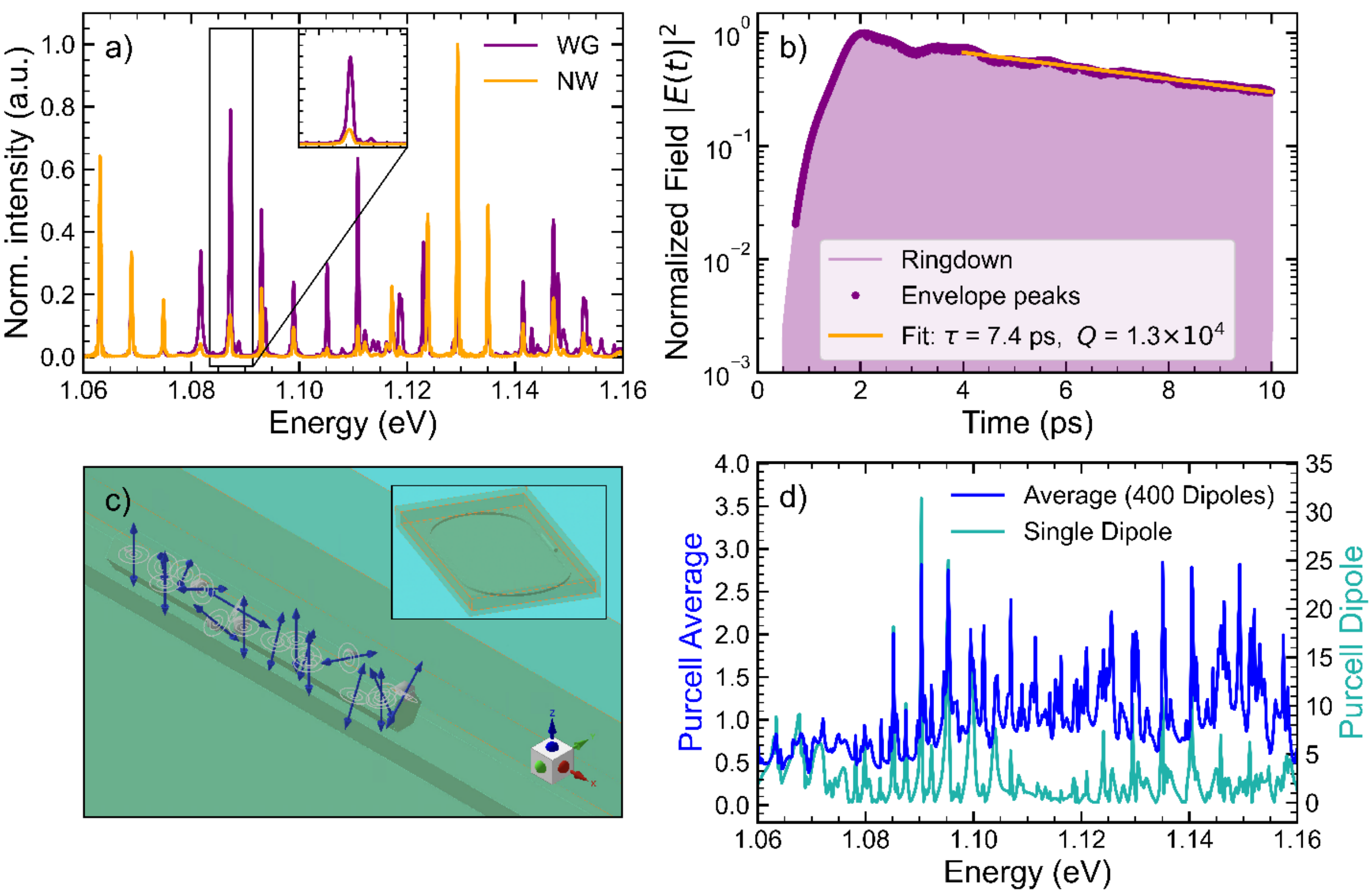

**Figure S3: FDTD analysis of NW-WG coupling and cavity enhancement.** (a) Fourier-transformed spectra recorded at time monitors inside the nanowire and silicon waveguide under $HE_{1,1,b}$ mode excitation. (b) Ringdown analysis of a selected resonance used to extract photon lifetime and cavity Q-factor. (c) FDTD simulation geometry for Purcell-factor calculations using 400 randomly distributed and oriented dipole emitters inside the NW (few dipoles shown for reference). (d) Spectrally resolved Purcell enhancement for the ensemble-averaged dipoles and for a single optimally positioned dipole.

**Figure S3** summarizes the FDTD simulations used to quantify optical coupling between the GaAsSb nanowire and the silicon racetrack resonator, as well as the resulting cavity-induced enhancement of spontaneous emission. In **Figure S3**a, the hybrid structure is excited by injecting the $HE_{1,1,b}$ -like nanowire mode. Time monitors are placed inside the NW and in the silicon waveguide (WG), and the recorded time-domain fields are Fourier transformed to obtain the corresponding spectra. Distinct resonances appear at identical energies in both the NW (orange) and WG (purple) spectra. This simultaneous appearance confirms that the resonant modes are not confined to only one component but are hybridized across the NW–WG system, demonstrating strong spatial overlap and efficient optical coupling. The inset emphasizes a representative resonance, showing that the spectral feature is clearly present in both monitoring regions. To quantify the resonator performance, **Figure S3**b presents a cavity ringdown analysis for the resonance highlighted in the inset of panel (a). After excitation, the temporal decay of the normalized electric field intensity $|E(t)|^2$ is evaluated. The envelope maxima are extracted (purple markers) and fitted with an exponential decay, appearing linear on the semilogarithmic scale (orange fit curve). From this fit, a photon lifetime of $\tau = 7.4$ ps is obtained, corresponding to a calculated quality factor of $Q = 1.3 \times 10^4$. This confirms that the hybrid system supports high-Q resonances enabling the observed feedback-assisted lasing. While the analysis in (a,b) probes coupling under guided-mode excitation, **Figure S3**c investigates the enhancement of spontaneous emission into the resonator modes. Here, the nanowire is excited using a set of randomly oriented dipole point sources distributed uniformly throughout the entire nanowire volume. Each dipole is assigned a random phase and orientation to mimic an incoherent ensemble emitter distribution. A total of 400 dipoles is used in the simulation (only a small number is shown schematically for clarity). This approach provides a realistic estimate of the average emission enhancement expected from the full gain volume. The resulting Purcell-factor calculations are shown in **Fig. S3**d. The dark-blue curve represents the ensemble-averaged Purcell factor obtained by averaging over all 400 dipoles (left axis). In addition, the light-blue curve shows a single dipole located at an optimal position and orientation, exhibiting exceptionally high Purcell enhancement up to 32 (right axis) at comparable energies with respect to our measured dominant lasing emission. Importantly, the Purcell enhancement directly impacts the emission coupling efficiency into the resonant mode. Since the spontaneous emission coupling factor is given by

$$\beta \approx \frac{F_P}{1 + F_P}$$

large Purcell factors lead to high $\beta$ factors. This implies that a significant fraction of spontaneous emission is funneled into the lasing modes, which increases mode competition and reduces the sharpness of the classical "S-shaped" input–output curve in nanolasers.[2] This provides a consistent explanation for the comparatively smooth lasing transition observed experimentally in such resonantly enhanced nanowire–cavity systems.

## S4. Micro-Photoluminescence (µ-PL) Spectroscopy

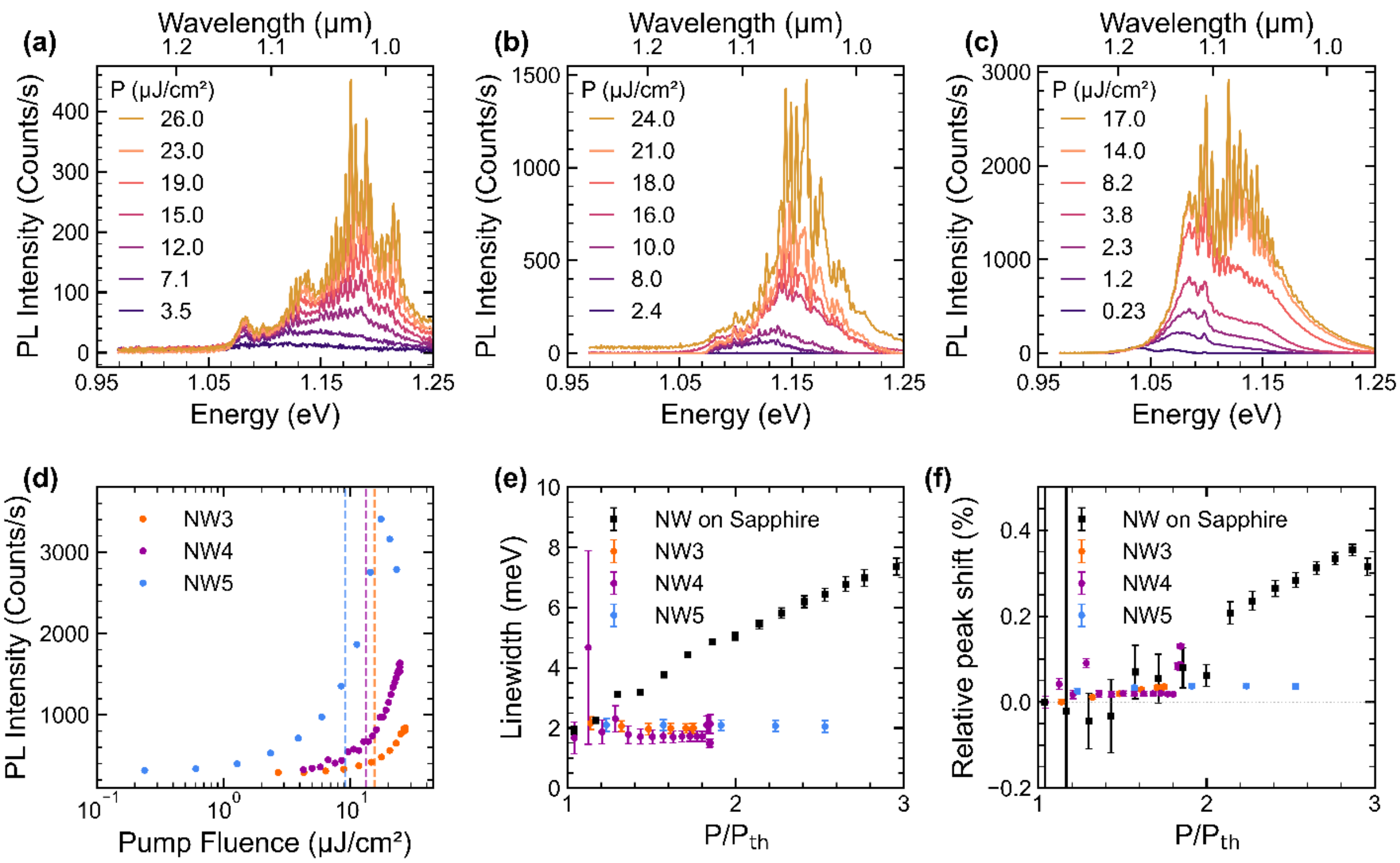

**Figure S4: Statistical fluence-dependent lasing characteristics of resonator-integrated nanowires.** (a–c) Pump-fluence dependent µ-PL spectra of three additional GaAsSb nanowires (NW3–NW5) transfer-printed onto silicon racetrack resonators. (d) Corresponding input–output characteristics indicating lasing thresholds in the range of 9.2 – 15.5 µJ/cm². (e) Emission linewidth evolution above threshold compared to the standalone nanowire on sapphire. (f) Relative shift of the dominant lasing peak above threshold, demonstrating strongly improved wavelength stability for resonator-coupled nanowires.

**Figure S4** provides additional device statistics demonstrating that the lasing characteristics reported in the main text are reproducibly obtained across multiple transfer-printed GaAsSb nanowires integrated on silicon racetrack resonators. Panels **(a–c)** show pump-fluence dependent µ-PL spectra for three further devices (NW3–NW5), each aligned and placed onto the straight arm of the racetrack resonator using the deterministic transfer-printing approach described in **Figure S1**b. For all nanowires, a clear transition from broadband spontaneous emission to narrow, sharply defined spectral features is observed upon increasing excitation fluence, consistent with the onset of laser oscillation. The corresponding input–output characteristics are summarized in **Figure S4**d. All three devices exhibit pronounced threshold behavior, confirming lasing action across the sample set. Extracted lasing thresholds range from 9.2 to 15.5 $\mu J/cm^2$, reflecting moderate device-to-device variations that are expected due to small differences in nanowire diameterand positioning.

Beyond threshold, the linewidth evolution shown in **Figure S4**e further highlights the stabilizing role of the passive silicon cavity. While the standalone reference nanowire on sapphire exhibits strong linewidth broadening with increasing pump fluence, all resonator-integrated devices maintain a nearly constant linewidth of approximately 2 meV over the investigated excitation range. This behavior indicates that the dominant lasing mode remains spectrally well-defined even at elevated carrier densities, in stark contrast to the strong broadening typically observed for nanowire Fabry–Pérot cavities where the active medium itself defines the resonator. Consistently, the enhanced frequency stability is further quantified in **Figure S4**f, which plots the relative shift of the dominant lasing peak above threshold with respect to its spectral position at $P=P_{th}$. For all three resonator-coupled nanowires, the peak position remains remarkably stable, with relative shifts below 0.05%, even at high pump levels. This is significantly smaller than the ~0.4% shift observed for the standalone nanowire on sapphire under comparable excitation conditions. The improved stability arises because the effective cavity is largely determined by the passive silicon racetrack resonator, thereby strongly reducing sensitivity to carrier-induced refractive index changes within the nanowire gain medium. Together, the statistical analysis confirms that resonator integration consistently yields nanowire lasers with improved spectral robustness, demonstrating the reproducibility of the hybrid NW-resonator platform.